\documentclass[a4paper,11pt,twoside]{article}
\usepackage{a4wide}

\newcommand{\beq}{\begin{equation}}
\newcommand{\eeq}{\end{equation}}
\newcommand{\beqa}{\begin{eqnarray}}
\newcommand{\eeqa}{\end{eqnarray}}
\newcommand{\ben}{\begin{enumerate}}
\newcommand{\een}{\end{enumerate}}
\newcommand{\bi}{\begin{itemize}}
\newcommand{\ei}{\end{itemize}}

\title{On the Financial Crisis 2008 from a Physicist's  Viewpoint: \newline 
A Spin-Glass Interpretation}

\author{U.\ Krey\footnote{e-mail uwe.krey@physik.uni-regensburg.de }
\\
  Inst.\ f\"ur Physik II, Universit\"at Regensburg, 93040 Regensburg,
Germany
  }
\date{Dec. 17, 2008; updated version as of Jan. 14, 2009}
\begin{document}

\maketitle
\begin{abstract}

\noindent In an informal way, a number of thoughts on the
financial crisis 2008 are presented from a physicist's viewpoint,
considering the problem as a nonergodicity transition of a spin-glass-like
system. Some suggestions concerning the way out of the
crisis are also discussed, concerning Keynesian {\it deficit spending}
methods, tax reductions, and finally the method {\it ruin and
recreate}, which is known from optimization theory. The de Almeida-Thouless
instabilty line of spin glass theory is also given a financial
interpretation.\end{abstract} {

\vglue 0.2 truecm\hrule\vglue 0.5 truecm

\section{Introduction} This is an informal communication, not intended for
publication. The topic, the financial crisis 2008 and its probable expansion
into a serious economic crisis, is closely connected to spin-glass physics,
(not only) to the opinion of the author. This is more or less common
wisdom of the {\it econophysics} community (\cite{Stauffer}). (If you
want to comment on some issue, please send mail).

Furthermore, the degrees of freedom of the system, which may be represented
by a two-dimensional graph, are, in a strong simplification, replaced by
binary (i.e., `Ising') degrees of freedom, $s_i$, where the signs, $\pm 1$,
may correspond to gains and losses, respectively, of the individual
companies competing for profit, which are represented by the vertices $i$ of
the graph and interact with each other in a global market.

These interactions are {\it frustrated} - a notion known from spin-glass
theory, e.g. around a closed loop with an odd number of edges, taking the
products (`Wilson loop products') of an odd number of the Ising variables,
the interactions - whatever they may be - will lead to a positive outcome at
the end, if the loop product is taken in one direction, and a negative
outcome, if it is in the opposite direction.  (The {\it frustration} or {\it
complexity} of the system will become important below.)

This possibility of inherent "frustration" in the loop of simultaneous
cooperation and competition, \cite{REM}, (a certain quenched correlation
of the system, leading to essentially unavoidable and  equally probable
gains and losses, where the probability of a {\it loss}\, in a "betting
situation" seems to have sometimes been forgotten by the global
"players", or shifted to future generations), is fixed by the
interactions on the graph, which is typical for a spin-glass system,
assuming that the graph itself is disordered, with fixed random
positions and interactions ("quenching").

Here is the place, where the interest rate, $r$, of the central bank of
the involved currency,  comes  into play. Since very low values
of $r$ enhance the tendency of the investment banks to invent new
"structured financial products", whatsoever, an enhancement(!) of the
interest rate should also be considered as a means to reduce the {\it
complexity} \, of the financial market, not only as a means to reduce
the danger of inflation. Of course, the primary effect of a change of $r$ is
in conflict with this consideration: primarily, at low values of $r$
everyone gets loans (e.g., for houses), which then become too expensive,
if $r$ is increasing again.

One should also consider that one is not dealing with a closed system, but
rather a system with two "sinks", the Iraque, and Afghanistan.

{\it "Whatever the interactions may be"}: this is a crucial text item.
Typically, since there are many "groups" ($N\gg 1$), the interactions of the
system may roughly correspond to a seemingly rather special spherical spin
glass, a "p-spin-glass", i.e. with p interacting spins, where $p\gg 1$ (many
companies, $p$, form an interacting group) and where the spherical condition
$\sum_{i=1}^N \,s_i^2\stackrel{!}{=}N$ is natural.

\section{More analogies}

The following remarks are obvious:

The interaction is described by a global optimization function,
corresponding physically to the {\it Hamiltonian} of the system, and is {\it
not}\, essential, in contrast to the degrees of freedom of the system and
some relevant macroscopic variables, e.g. the {\it specific volume} or {\it
specific size} $v$ of a company or group of cooperating companies. In an
{\it ideal situation} (remember {\it the ideal gas}) one thinks '{\it The
larger the better}', because of the {\it synergies}, but the {\it reality}
may be different: recently one has apparently come to the opposite
conclusion that '{\it As small as possible}' is more beautiful, because in
this way the costs seem to be reduced more effectively (unfortunately, to
say it mildly, in both cases the unemployment is enhanced.) This change of
paradigm - the analogon is a change of slope of the saturation pressure
$p(v)$ - may be considered as a warning signal for criticality.

 Another essential physical parameter, the temperature, seems to correspond
to the {\it economic activity}: enhancing (or reducing) the temperature,
respectively, is perhaps analogous to turning up (or slowing down) the
economic activity (the analogy is more visible in the German language:
"Temperatur" {\it versus} "Konjunktur"). Finally, the physical variable $p$
({\it pressure}) may be translated economically into a {\it "reciprocal
specific wealth"} or {\it "individual economic pressure"}, an intensive
quantity, in constrast to the size parameter $V$, which is extensive.
($v=\frac{V}{N}$ is the {\it specific company size, or specific group size},
where $N$ is the number of companies or groups of cooperating companies).

 Furthermore, physically there are two "fluid" phases of a system, a
{\it vapor phase}, which is a {\it wealthy} phase in our economical
interpretation, and a {\it liquid} (or  {\it poor}) phase, respectively.

Here the well-known {\it Van der Waals theory} (see e.g. \cite{Krey}) of the
{\it vapor $\Leftrightarrow$ liquid transition} is set into an economical
analogy between economic phases with two different degrees of {\it wealth}.
Physically one has a {\it stable} and a {\it metastable} $vapor$ phase and
two corresponding $liquid$ phases, which are separated from the $vapor$ by
the famous {\it Maxwell line}, which is reached at a value $v_{\,\,\rm
|Maxwell}$, separating also the ranges of stability and metastability.
Furthermore, if one starts in the "wealthy" phase and reduces $v$ gradually
below $v_{\,\,\rm |Maxwell}$, there is a critical size $v_c < v_{\,\,\rm
|Maxwell} $, where with gradually decreasing $v$ the metatastable system
becomes unstable and jumps to the "{\it poor}" phase.

\section{Non-ergodicity}

In the p-spin-glass analogy (see e.g. \cite{Vulpiani}) the instability is a
dynamic {\it freezing}\, phenomenon, which may correspond to G\"otze's
well-known mode-coupling theory, \cite{Goetze}, which descibes a kind of
{\it transition to immobility}\, in the context of amorphization of liquid
metals. Economically, the nonergodicity means, for example, that the
financial institutions do no longer cooperate with each other, nor with
their clients. To get out of the crisis, there are more-or-less obvious
prescriptions: the most direct way is to follow the well-known {\it
Keynesian} measures of enhancing the "Konjunktur" by {\it deficit spending}
\, of the state. Physically this corresponds to an enhancement of the
temperature, i.e. a {\it vertical} perturbation in the usual axes. In
contrast, a {\it tax reduction}, which is also under discussion, would
correspond to a more {\it horizontal} perturbation, which seems less
effective (of course, computer simulations on this issue might be useful).
One may also try to reduce the "frustration" possibility (or {\it
complexity}) of the system, by forbidding extremely risky financial
products, e.g., the possibility of {\it short selling} leased assets.

\section{Replica-symmetry breaking}

The signature of spin-glass behaviour, and specifically of the related
non-ergodicity transition, is 'replica-symmetry breaking', which is a
complicated issue, which we don't treat - in an elementary way, one might
say that there are unsurmountable barriers between the ergodic components.
Details can again be found in \cite{Vulpiani}. Especially, it should be
noted that spherical p-spin-glasses, which are here under consideration,
have only 1-step replica symmetry breaking, i.e., there are only two
hierarchies, may be the "normal" financial products, and the "structured
financial derivates" (in any case, the two fluid states of the Van-der-Waals
theory should be related). Actually, although there the risk is broadened
both in space and also in time, by some kind of "interassuring", the
promises concerned are {\it higher gains}, which also means {\it higher
losses}\, in case of a failure since the probability distributions involved
are usually symmetric and have "fat tails" both on the gain and on the loss
sides. 

The essential point is the danger that the system has become eventually so
complex that the companies "assuring the risk"  belong to different,
noncooperating 'valleys' of the spin-glass system, after non-ergodicity has
appeared.

\section{Ways out of the crisis}

A combination of the above-mentioned methods may be useful, since the
optimal perturbation is  perhaps essentially vertical, but not exactly.
Numerical simulations on this issue would be helpful. In an extreme case
one might also consider brute-force methods, as the method {\it ruin and
recreate}, which is known from optimization theory, and is actually not
so brute-force as it sounds, but rather smooth, \cite{Schneider}.

According to a hint of T. Preis, it may be essential to take into account
that financial
processes are {\it nonstationary}. But even then the spherical condition,
$\sum_{i=1}^N s_i^2=N$, might remain valid, as well as the interpretation
of the crisis as a freezing into a nonergodic state of a p-state-spin-glass
model as described by \cite{Vulpiani}.

\section{Countermeasures of the state and the de Almeida-Thouless line}
The capital flow from the state as a countermeasure against the crisis
could be analogous to the external field $h$ of the spin-glass theory.
As a consequence, the famous de Almeida-Thouless instability line, 
\cite{AT}, to
surmount the crisis, should follow the law $h^{2/3}\propto T_f\,,$
where $T_f$ is the freezing temperature ("Konjunktur") of the system.

\section*{Acknowledgement}
The author thanks T. Preis for useful remarks.

\end{document}